\documentclass[a4paper,11pt]{article}

\usepackage{latexsym}
\usepackage{a4wide}
\usepackage{graphicx, subfigure}
\usepackage{cite}
\usepackage{tabularx}
\usepackage{array}
\usepackage{graphics}
\usepackage{amsmath}
\usepackage{amssymb}
\usepackage{longtable}
\usepackage{verbatim}
\usepackage[table]{xcolor}
\usepackage{booktabs}

\definecolor{light-gray}{gray}{0.90}

\renewcommand{\arraystretch}{1.2} 

\newcommand{\cen}[1]{\multicolumn{1}{c}{#1}}
\newcommand{\ra}[1]{\renewcommand{\arraystretch}{#1}}
\newcommand{\rb}[1]{\renewcommand{\tabcolsep}{#1}}
\newcommand{\av}[1]{\langle #1 \rangle}

\begin{document}

\hfill {\tt CERN-PH-TH/2014-198,   MITP/14-078}

\def\thefootnote{\fnsymbol{footnote}}

\begin{center}

\vspace{3.cm}

{\huge\bf Global fits to $b \to s \ell\ell$ data and signs for lepton non--universality}

\vspace{2.cm}
{\Large\bf  
T.~Hurth\footnote{Email: tobias.hurth@cern.ch}$^{,a}$,
F.~Mahmoudi\footnote{Also Institut Universitaire de France, 103 boulevard Saint-Michel, 75005 Paris, France\\ \hspace*{0.49cm} Email: nazila@cern.ch}$^{,b,c}$,
S.~Neshatpour\footnote{Email: neshatpour@ipm.ir }$^{,d}$
}

\vspace{1.cm}
{\em $^a$PRISMA Cluster of Excellence and  Institute for Physics (THEP)\\
Johannes Gutenberg University, D-55099 Mainz, Germany}\\[0.2cm]
{\em $^b$Universit{\' e} de Lyon, Universit{\' e} Lyon 1, F-69622 Villeurbanne Cedex, France;\\
Centre de Recherche Astrophysique de Lyon, Saint-Genis Laval Cedex, F-69561, France; CNRS, UMR 5574;
Ecole Normale Sup{\' e}rieure de Lyon, France}\\[0.2cm]
{\em $^c$Theory Division, CERN, CH-1211 Geneva 23, Switzerland}\\[0.2cm]
{\em $^d$School of Particles and Accelerators,
Institute for Research in Fundamental Sciences (IPM)
P.O. Box 19395-5531, Tehran, Iran}

\end{center}

\renewcommand{\thefootnote}{\arabic{footnote}}
\setcounter{footnote}{0}

\vspace{1.cm}
\thispagestyle{empty}
\centerline{\bf ABSTRACT}
\vspace{0.5cm}

There are some slight tensions with the SM predictions within the latest LHCb measurements. Besides the known anomaly in one angular observable of the rare decay $B \rightarrow K^{*} \mu^+\mu^-$, another small discrepancy recently occurred. The ratio $R_K = {\rm BR}(B^+ \to K^+ \mu^+ \mu^-) / {\rm BR}(B^+ \to K^+ e^+ e^-)$ in the low-$q^2$ region has been measured by LHCb showing a $2.6\sigma$ deviation from the SM prediction. In contrast to the anomaly in the rare decay $B \to K^{*} \mu^+\mu^-$ which is affected by power corrections, the ratio  $R_K$ is theoretically rather clean. 
We analyse all the $b \to s \ell\ell$ data with global fits and in particular explore the possibility of breaking of lepton universality. 
Possible cross-checks with an analysis of the inclusive $B \to X_s \ell^+\ell^-$ decay are also explored.
 
\newpage

\section{Introduction}

There are some small tensions with the SM predictions within the latest LHCb measurements:
The first measurement of new angular observables in the exclusive decay $B \to K^* \mu^+\mu^-$ has shown  a kind of anomaly~\cite{Aaij:2013qta}.  Due to the large hadronic uncertainties  it is not clear whether this anomaly is a first sign for new physics beyond the SM, or a consequence of underestimated hadronic power corrections~\cite{Jager:2012uw,Descotes-Genon:2013wba,Altmannshofer:2013foa,Hambrock:2013zya,Gauld:2013qba,Buras:2013qja,Gauld:2013qja,Datta:2013kja,Beaujean:2013soa,Horgan:2013pva,Buras:2013dea,Hurth:2013ssa,Descotes-Genon:2014uoa,Khodjamirian:2010vf,Lyon:2014hpa}; but of course, it could just turn out being a statistical fluctuation. So the LHCb analysis based on the 3 fb$^{-1}$ dataset is eagerly awaited to clarify the situation.

Besides this known  anomaly in the angular analysis of the rare decay $B \rightarrow K^{*} \mu^+\mu^-$, another small discrepancy 
recently occurred. The ratio $R_K = {\rm BR}(B^+ \to K^+ \mu^+ \mu^-) / $ ${\rm BR}(B^+ \to K^+ e^+ e^-)$ in the low-$q^2$ region has been measured by LHCb using the full 3 fb$^{-1}$ of data, showing a $2.6\sigma$ deviation from the SM prediction~\cite{Aaij:2014ora}. 
In contrast to the anomaly in the rare decay $B \rightarrow K^{*} \mu^+\mu^-$ which is affected by unknown power corrections, the ratio  $R_K$ is theoretically rather clean. 

The $R_K$ discrepancy has been addressed in a few recent studies~\cite{Alonso:2014csa,Hiller:2014yaa,Ghosh:2014awa,Biswas:2014gga,Straub}.
In Ref.~\cite{Hiller:2014yaa}, two leptoquark models have been proposed to explain the discrepancy. In Ref.~\cite{Biswas:2014gga}, $R$-parity violating supersymmetry is used to explain the $R_K$ anomaly together with lepton non-universal effects in the $W_R$ search observed by CMS.
The authors of Ref.~\cite{Ghosh:2014awa} have studied the $R_K$ result by performing a Bayesian statistical fit to the Wilson coefficients considering the data on some $b\to s \ell \ell$ transitions.

In this paper, we analyse the latest LHCb data within various global fits applying a frequentist statistical approach using the available data on all the relevant $|\Delta B|= |\Delta S| =1$ processes in order to explore the $R_K$ and $B\to K^* \mu^+ \mu^-$ anomalies.

Moreover, we discuss the inclusive decay mode $B \to X_s \ell^+\ell^-$ which is an important theoretically clean mode of the indirect search for new physics via flavour observables~\cite{Hurth:2007xa,Hurth:2003vb}; especially it allows for a non-trivial cross-check of the recent LHCb data on the exclusive mode~\cite{Hurth:2013ssa}.

This paper is organised as follows. In section 2 we describe our model independent analysis and highlight the main new theoretical and experimental inputs of the present study. In section 3 we present our results for the global fits for different sets of operators. The comparison with the inclusive decay mode is provided in section 4 and our conclusions are given in section 5.

\section{Input of model-independent analysis}

Compared to the analysis in Ref.~\cite{Hurth:2013ssa}, we have the following changes within the experimental and theoretical inputs: 

\begin{itemize}

\item We now also consider the recent LHCb measurements of $B \rightarrow K \ell^+\ell^-$ and in particular the observable $R_K$~\cite{Hiller:2003js}.
Our theory analysis in the low-$q^2$ region is based on the method of QCD factorisation (QCDf). The factorisable and non-factorisable order one $\alpha_s$ corrections are known from Refs.\cite{Charles:1998dr,Beneke:2000wa, Beneke:2001at,Beneke:2004dp}.
In our numerical analysis we have considered the aforementioned corrections following~\cite{Bobeth:2007dw}.
We have chosen the factorisation scheme according to~\cite{Beneke:2000wa} where one of the three independent $B \to K$ form factors, $f_+$, is fixed by $f_+\equiv \xi_K$ to all orders in perturbation theory. This  $f_+$ form factor is taken from LCSR calculations of Ref.~\cite{Khodjamirian:2010vf}.
For the high-$q^2$ region we have followed Ref.~\cite{Bobeth:2011nj}, where instead of using LCSR extrapolated form factors and applying Isgur-Wise relation, we have used the three full form factors from lattice calculations~\cite{Bouchard:2013eph} which significantly reduces the form factor uncertainties.

\item For the $B \rightarrow K^{*} \mu^+\mu^-$ angular observables, we use the lattice results for the form factors of Ref.\cite{Horgan:2013hoa} in the high-$q^2$ region thereby getting a significant reduction in the theoretical uncertainties.
The theoretical uncertainties due to power corrections are estimated following Refs.~\cite{Egede:2008uy,Egede:2010zc}.  
In order to be conservative in our theoretical error estimation,  we have doubled the power correction budget in the overall error compared to the theoretical
predictions in Ref.~\cite{Descotes-Genon:2013vna}.

\item The three loop QCD corrections for $C_{10}$~\cite{Hermann:2013kca} as well as the NLO electroweak corrections~\cite{Bobeth:2013tba} are included.

\item We also consider the exclusive electron $B \rightarrow K^{*} e^+e^-$ decay for the region where 
$q^2$ takes the whole $[0.1,(M_B -M_{K^*})^2]\;{\rm GeV}^2$ range.\footnote{Although it seems that 
using the experimental results of 
LHCb~\cite{Aaij:2013hha} ($(3.1 \pm 0.9)\times 10^{-7}$)  available for 
the $[0.0009,1.0]\;{\rm GeV}^2$ bin and its corresponding SM prediction ($(2.43 \pm 0.57)\times 10^{-7}$)~\cite{Jager:2012uw} 
one could get an observable with less (theoretical + experimental) error, but we use the full range of $q^2$ from BaBar since the 
$[0.0009,1.0]\;{\rm GeV}^2$ bin is dominated by $C_7^{\rm eff}$.}

\item For the inclusive modes $B \to X_s \ell^+ \ell^-$ we consider the electron and muon final states separately and for the experimental
values we use the BaBar results~\cite{Lees:2013nxa}.

\item We have updated all the numerical values for the input parameters~\cite{Aoki:2013ldr,Agashe:2014kda}. Some of the main input parameters which have been updated compared to~\cite{Hurth:2013ssa} are given in Table~\ref{tab:input}.

\item Compared to Refs.~\cite{Hurth:2013ssa,Mahmoudi:2014mja}, where there is an 11\% theoretical uncertainty for the SM prediction of 
BR($B_s \to \mu^+ \mu^-$), we now have an overall error of about 7.5\%. This reduction is due to the improved precision
of the lattice results for the decay constant $f_{B_s}$~\cite{Aoki:2013ldr} as well as inclusion of higher order QCD and EW corrections\footnote{
See Ref.~\cite{Mahmoudi:2012un} for the various sources of uncertainties.}.

\end{itemize}

\begin{table}[!t]
\begin{center}
\footnotesize{\begin{tabular}{|lr|lr|}\hline
$M_{B^0}=5.27958 (17)$ GeV                  & \cite{Agashe:2014kda} & $M_{K^*}=0.89166 (26)$ GeV                & \cite{Agashe:2014kda} \\
$M_{B^+}=5.27926 (17)$ GeV                  & \cite{Agashe:2014kda} & $M_{K^0}=0.497614 (24)$ GeV               & \cite{Agashe:2014kda} \\ \hline
$ f_B=190.5 \pm 4.2 $ MeV        & \cite{Aoki:2013ldr}            & $ \tau_{B_s} = 1.512 \pm 0.007\ {\rm ps}$ & \cite{Agashe:2014kda} \\
$ f_{B_s} = 227.7 \pm  4.5$ MeV  & \cite{Aoki:2013ldr}            &                                           &                       \\ \hline
$a_{1}^{K}$(1 GeV)$=0.06\pm0.03$            & \cite{Ball:2006fz}    & $f_K=156 \pm 5$ MeV                   & \cite{Bazavov:2009bb} \\
$a_{2}^{K}$(1 GeV)$=0.25 \pm0.15$           & \cite{Ball:2006wn}    &                                           &                       \\ \hline
\end{tabular}}
\caption{Input parameters. \label{tab:input}}
\end{center}
\end{table}

The list of all the observables that we have considered in this work together with the SM values and the experimental results are given in Tables~\ref{tab:BtoKstar} and \ref{tab:rest}. The theoretical predictions of all the observables are computed using the program {\tt SuperIso}~\cite{Mahmoudi:2007vz,Mahmoudi:2008tp}. For more details
on the theoretical framework we refer to Ref.~\cite{Hurth:2013ssa,Hurth:2012jn}. 

We perform a model independent $\chi^2$ analysis using all the observables given in Tables~\ref{tab:BtoKstar} and \ref{tab:rest}. For $B \rightarrow K^{*0} \mu^+\mu^-$ angular observables we consider the experimental correlations as described in Ref.~\cite{Hurth:2013ssa}. As one of the main objectives of this study is to investigate lepton universality and assess the effect of $R_K$, we use the $\Delta \chi^2$ approach which is more suitable to find the preferred directions in the new physics parameter space. 
But we have to check first that the  $\chi^2$ method signals the overall consistency of the fit. 

For each Wilson coefficient, one defines $\delta C_i = C_i^{\rm NP} - C_i^{\rm SM}$. We consider separately the electron and muon semileptonic Wilson coefficients $C_{9,10}^e$ and $C_{9,10}^\mu$ respectively, which are equal in the SM or in models with lepton universality. 

\begin{table}
\ra{0.90}
\rb{1.3mm}
\begin{center}
\rowcolors{1}{}{light-gray}
\footnotesize{\begin{tabular}{@{}lrrc||lrr@{}}
\toprule[.6pt]
Observable & \cen{SM prediction} & Measurement        &  & Observable & \cen{SM prediction} & Measurement \\ 
\hline \hline
%
\rowcolor{yellow}$q^2 \in [\,0.1\,,\,2\,]\,{\rm GeV}^2 $& & & &  $q^2 \in [\,14.18\,,\,16.0\,]\,{\rm GeV}^2 $ & &    \\  
$\av{P_1}$  & $0.01  \pm{0.06} $ & $-0.19 \pm 0.40$   & & $\av{P_1}$  & $-0.49 \pm{0.41} $ & $0.07 \pm 0.28$   \\ 
$\av{P_2}$  & $0.16  \pm{0.03} $ & $0.03 \pm 0.15$    & & $\av{P_2}$  & $-0.43 \pm{0.04} $ & $-0.50 \pm 0.03$ \\ 
$\av{P'_4}$ & $-0.39 \pm{0.05} $ & $0.00 \pm 0.52$    & & $\av{P'_4}$ & $1.22 \pm{0.14} $ & $-0.18 \pm 0.70$  \\ 
$\av{P'_5}$ & $0.51  \pm{0.05} $ & $0.45 \pm 0.24$    & & $\av{P'_5}$ & $-0.71 \pm{0.16} $ & $-0.79 \pm 0.27$  \\ 
$\av{P'_6}$ & $-0.06 \pm{0.04} $ & $0.24 \pm 0.23$    & & $\av{P'_6}$ & $0.00 \pm{0.00} $ & $0.18 \pm 0.25$  \\ 
$\av{P'_8}$ & $0.03  \pm{0.05} $ & $-0.12 \pm 0.56$   & & $\av{P'_8}$ & $0.00 \pm{0.02} $ & $-0.40 \pm 0.60$  \\ 
$\av{F_L}$  & $0.31  \pm{0.18} $ & $0.37 \pm 0.11$    & & $\av{F_L}$  & $0.31 \pm{0.08} $ & $0.33 \pm 0.09$  \\ 
$10^7\,{\rm GeV}^2 \times \av{d{\rm BR}/dq^2}$  & $0.95 \pm{0.70}$ & $0.60 \pm 0.10$ & & $10^7\,{\rm GeV}^2 \times \av{d{\rm BR}/dq^2}$  & $0.75 \pm{0.43} $ & $0.56 \pm 0.10$  \\  %
\rowcolor{white} & &    \\ [-12pt] 
\midrule[.5pt]
\rowcolor{yellow}$q^2 \in [\,2.0\,,\,4.3\,]\,{\rm GeV}^2 $& & & &  $q^2 \in [\,16.0\,,\,19.0\,]\,{\rm GeV}^2 $ &   &    \\  
$\av{P_1}$  & $-0.05 \pm{0.09} $ & $-0.29 \pm 0.65$ & & $\av{P_1}$  & $-0.69 \pm{0.26} $ & $-0.71 \pm 0.36$\\ 
$\av{P_2}$  & $0.26  \pm{0.09} $ & $0.50 \pm 0.08$  & & $\av{P_2}$  & $-0.35 \pm{0.07} $  & $-0.32 \pm 0.08$\\ 
$\av{P'_4}$ & $0.51  \pm{0.07} $ & $0.74 \pm 0.60$  & & $\av{P'_4}$ & $1.30 \pm{0.10} $ & $0.70 \pm 0.52$ \\ 
$\av{P'_5}$ & $-0.35 \pm{0.11} $ & $0.29 \pm 0.40$  & & $\av{P'_5}$ & $-0.54 \pm{0.15} $ & $-0.60 \pm 0.21$ \\ 
$\av{P'_6}$ & $-0.07 \pm{0.05} $ & $-0.15 \pm0.38$  & & $\av{P'_6}$ & $0.00 \pm{0.00} $ & $-0.31 \pm 0.39$ \\ 
$\av{P'_8}$ & $0.06  \pm{0.05} $ & $-0.3 \pm 0.60$  & & $\av{P'_8}$ & $0.00 \pm{0.01} $ & $0.12 \pm 0.54$ \\ 
$\av{F_L}$  & $0.76  \pm{0.17} $ & $0.74 \pm 0.10$  & & $\av{F_L}$  & $0.30 \pm{0.04} $ & $0.38 \pm 0.09$ \\ 
$10^7\,{\rm GeV}^2 \times \av{d{\rm BR}/dq^2}$  & $0.45 \pm{0.25} $ & $0.30 \pm 0.05$ & & $10^7\,{\rm GeV}^2 \times \av{d{\rm BR}/dq^2}$  & $0.57 \pm{0.32} $ & $0.41 \pm 0.07$ \\ %
\rowcolor{white} & &   \\ [-12pt] 
\midrule[.5pt]
\rowcolor{yellow} $q^2 \in [\,4.3\,,\,8.68\,]\,{\rm GeV}^2 $ & & \\ 
$\av{P_1}$  & $-0.11 \pm{0.11} $ & $0.36 \pm 0.31$   \\ 
$\av{P_2}$  & $-0.38 \pm{0.04} $ & $-0.25 \pm 0.08$ \\ 
$\av{P'_4}$ & $0.99  \pm{0.06} $ & $1.18 \pm 0.32$   \\ 
$\av{P'_5}$ & $-0.85 \pm{0.07} $ & $-0.19 \pm 0.16$  \\ 
$\av{P'_6}$ & $-0.03 \pm{0.11} $ & $0.04 \pm 0.16$    \\ 
$\av{P'_8}$ & $0.02  \pm{0.13} $ & $0.58 \pm 0.38$   \\
$\av{F_L}$  & $0.64  \pm{0.20} $ & $0.57 \pm 0.08$   \\ 
$10^7\,{\rm GeV}^2 \times \av{d{\rm BR}/dq^2}$  & $0.61 \pm{0.38} $ & $0.49 \pm 0.08$   \\ %
\end{tabular} }
\caption{The SM predictions and experimental values of the $B\to K^* \mu^+ \mu^-$ observables used in this study. 
The experimental values are from~\cite{Aaij:2013iag,Aaij:2013qta}, here the experimental errors have been symmetrised by 
taking the largest side error and whenever there have been several sources of uncertainty the total error has been obtained by adding them in quadrature.
\label{tab:BtoKstar}}
\end{center} 
\ra{1.2}
\begin{center}
\footnotesize{
\begin{tabular}{@{}lrrl@{}}
\toprule[1.1pt]
Observable & SM prediction  &  ~~~Measurement &    \\ [1mm]
\hline \hline
$10^9\times{\rm BR}(B_s \to \mu^+ \mu^-)$ & $ 3.54 \pm 0.27 $ & $2.9 \pm 0.7$ & \cite{Aaij:2013aka,Chatrchyan:2013bka,CMSandLHCbCollaborations:2013pla} \\ [1mm]
$10^{10}\times{\rm BR}(B_d \to \mu^+ \mu^-)$ & $ 1.07 \pm 0.27 $ & $3.6 \pm 1.6$ & \cite{Aaij:2013aka,Chatrchyan:2013bka,CMSandLHCbCollaborations:2013pla} \\ [1mm] \hline
$R_{K\:{q^2 \in [1.0,6.0]({\rm GeV})^2}}$ & $1.0006 \pm 0.0004$ & $0.745\pm{0.097}$ & \cite{Aaij:2014ora} \\ [1mm] \hline
$10^9\,{\rm GeV}^2 \times \av{d{\rm BR}/dq^2}\left( B^0 \to K^0 \mu^+ \mu^- \right)_{q^2 \in [\,1.1\,,\,6.0\,]\,{\rm GeV}^2}$  & $31.7 \pm{9.4}$  & $18.7 \pm{3.6}$ & \cite{Aaij:2014pli} \\ [1mm] 
$10^9\,{\rm GeV}^2 \times \av{d{\rm BR}/dq^2}\left( B^0 \to K^0 \mu^+ \mu^- \right)_{q^2 \in [\,15.0\,,\,22.0\,]\,{\rm GeV}^2} $  & $13.6 \pm{2.0}$  & $9.5 \pm{1.7}$ & \cite{Aaij:2014pli}  \\ [1mm] \hline
$10^9\,{\rm GeV}^2 \times \av{d{\rm BR}/dq^2}\left( B^+ \to K^+ \mu^+ \mu^- \right)_{q^2 \in [\,1.1\,,\,6.0\,]\,{\rm GeV}^2}$  & $34.8 \pm{10.3}$  & $24.2 \pm{1.4}$ & \cite{Aaij:2014pli}  \\ [1mm]
$10^9\,{\rm GeV}^2 \times \av{d{\rm BR}/dq^2}\left( B^+ \to K^+ \mu^+ \mu^- \right)_{q^2 \in [\,15.0\,,\,22.0\,]\,{\rm GeV}^2}$  & $14.8 \pm{2.0}$  & $12.1 \pm{0.7}$ & \cite{Aaij:2014pli} \\ [1mm] \hline
$10^9\,{\rm GeV}^2 \times \av{d{\rm BR}/dq^2}\left( B^+ \to K^{*+} \mu^+ \mu^- \right)_{q^2 \in [\,1.1\,,\,6.0\,]\,{\rm GeV}^2} $ & $50.5 \pm{28.6}$  & $36.6 \pm{8.7}$ & \cite{Aaij:2014pli} \\ [1mm]
$10^9\,{\rm GeV}^2 \times \av{d{\rm BR}/dq^2}\left( B^+ \to K^{*+} \mu^+ \mu^- \right)_{q^2 \in [\,15.0\,,\,19.0\,]\,{\rm GeV}^2} $ & $61.5 \pm{34.8}$  & $39.5 \pm{8.5}$ & \cite{Aaij:2014pli}  \\ [1mm] \hline
$10^{6}\times{\rm BR}\left( B \to K^* e^+ e^- \right)_{q^2 \in [\,0.1\,,\,(M_B-M_{K^*})^2\,]\,{\rm GeV}^2}$ & $1.24 \pm{0.80} $ & $1.03 \pm{0.19} $ & \cite{Agashe:2014kda}  \\ [1mm] \hline  
$10^{6}\times{\rm BR}\left( B\to X_s e^+ e^- \right)_{q^2 \in[\,1\,,\,6\,] \,{\rm GeV}^2}$     & $ 1.73  \pm 0.12 $ &  $1.93 \pm{0.55}$ & \cite{Lees:2013nxa} \\ [1mm]
$10^{6}\times{\rm BR}\left( B\to X_s e^+ e^- \right)_{q^2 > 14.2 \,{\rm GeV}^2}$        & $ 0.20 \pm 0.06 $ &  $0.56 \pm{0.19}$ & \cite{Lees:2013nxa} \\ [1mm] \hline 
$10^{6}\times{\rm BR}\left( B\to X_s \mu^+ \mu^- \right)_{q^2 \in[\,1\,,\,6\,] \,{\rm GeV}^2}$ & $ 1.66 \pm 0.12 $ &  $0.66 \pm{0.88}$ & \cite{Lees:2013nxa} \\ [1mm]
$10^{6}\times{\rm BR}\left( B\to X_s \mu^+ \mu^- \right)_{q^2 > 14.2 \,{\rm GeV}^2}$    & $ 0.24 \pm 0.07 $ &  $0.60 \pm{0.31}$ & \cite{Lees:2013nxa} \\ [1mm] \hline
\end{tabular} 
}
\caption{The SM predictions and experimental values of observables used in this study. 
The experimental errors have been symmetrised by taking the largest side error and whenever there have been several sources of uncertainty the total error has been obtained by adding them in quadrature.
The theoretical errors have been symmetrised by averaging.}\label{tab:rest}
\end{center} 
\end{table}  

\section{Results}

We first make a global fit to all observables considering new physics contributions to two Wilson coefficients only at a time. We then extend the study by considering new physics contributions to four Wilson coefficients and highlight the limitations of considering arbitrarily only a subset of those. By comparing different sets of two and four operators, we can assess the influence of the different operators.  

In the following we use all the observables given in Tables~\ref{tab:BtoKstar} and \ref{tab:rest} in the global fits. It was shown in Ref.~\cite{Alonso:2014csa,Hiller:2014yaa} that the current data on $R_K$ cannot be explained by tensor operators only. Moreover, the bounds from $B_s\to \mu^+\mu^-$ disfavour also the possibility of scalar and pseudoscalar operators accounting for $R_K$. However, a fine-tuned solution at the 2$\sigma$ level remains possible if one assumes large electron contribution and one accepts cancellations in order to fulfil the $B_s\to \mu^+\mu^-$ constraints. Therefore we consider here new physics contributions to the vector and axial vector operators allowing for flavour non-universality. 

\subsection{Fit results for two operators}
\label{sec:2-op}
We first consider new physics effects in $O_9$ and $O_{10}$ and make a $\chi^2$ fit by scanning over $\delta C_9$ and $\delta C_{10}$. The minimum $\chi^2$ here is 52, with 52 degrees of freedom. The best fit point has therefore a correct value with respect to the goodness-of-fit.
\begin{figure}[!h]
\begin{center}
\includegraphics[width=7.cm]{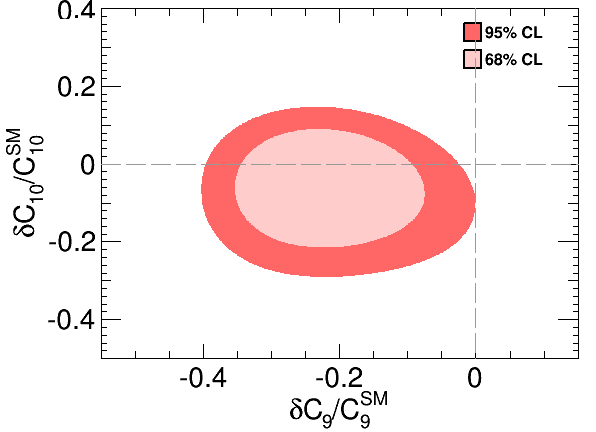}
\caption{Global fit results for $C_9, C_{10}$.}
\label{fig:2-c9_c10}
\end{center}
\end{figure}

The result is presented in Fig.~\ref{fig:2-c9_c10} where $\delta C_9$ and $\delta C_{10}$ are normalised to their SM values.
We notice that deviations of the order of 40\% compared to the SM predictions are allowed for $C_9$ and $C_{10}$ as a result of the global fit to all observables. However, the SM value (corresponding to $\delta C_9=\delta C_{10}=0$) is slightly disfavoured by 2.3$\sigma$ which represents a small tension for $C_9$. 

We then analyse further the effect of $C_9$ by considering separately the electron and muon contributions,
and make a $\chi^2$ fit by scanning over $\delta C_9^{\mu}, \delta C_9^{e}$. We obtain a minimum $\chi^2$ of 44, with 52 degrees of freedom, which shows that the fit is better than in the previous case. 

\begin{figure}[!h]
\begin{center}
\includegraphics[width=7.cm]{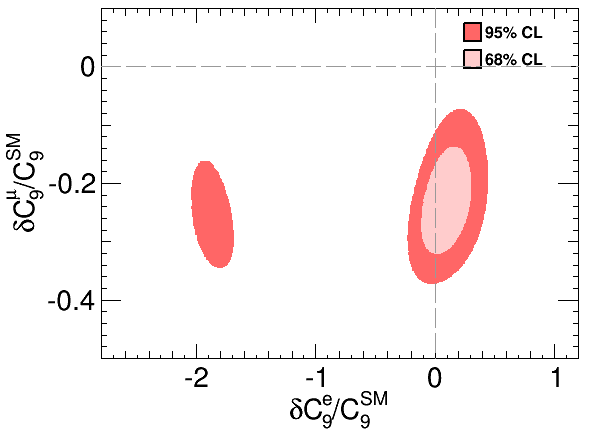}
\caption{Global fit results for $C_9^\mu, C_9^e$.}
\label{fig:2-c9e_c9m}
\end{center}
\end{figure}

Fig.~\ref{fig:2-c9e_c9m} shows the results in this two dimensional plane where $\delta C_9^{\mu}$ and $\delta C_9^{e}$ are normalised to their SM values. 
Two sets of solutions are found which manifest themselves in two separate zones in the figure. 
The most favoured one includes $\delta C^e_9=0$. Both zones incorporate similar values of $\delta C^\mu_9 \sim [-0.3,-0.1]\times C^{\rm SM}_9$, which shows that the SM value for the muon coefficient is disfavoured at more than 3$\sigma$. Yet, the universality condition, $\delta C^\mu_9=\delta C^e_9$, is still barely allowed, at the border of the 2$\sigma$ level, meaning that non-universality improves the fit.
This shows clearly that in this fit the tension with the SM originates from the muon contribution.

\begin{figure}[!t]
\begin{center}
\includegraphics[width=7.cm]{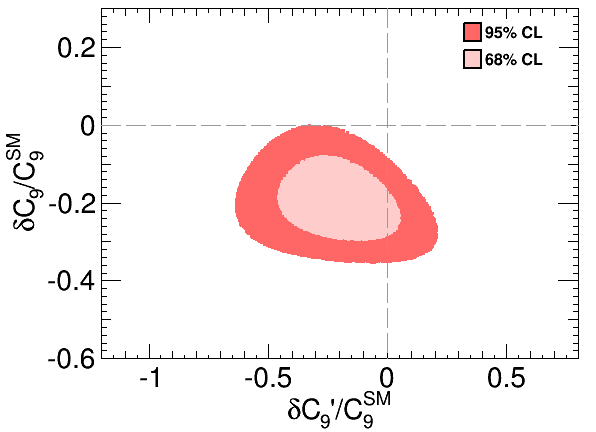}
\caption{Global fit results for $C_9, C_9^{'}$.}
\label{fig:2-c9_c9p}
\end{center}
\end{figure}

Finally we consider the possibility of chirality flipped operator $O_9^{'}$ and scan over $\delta C_9, \delta C_9^{'}$. The minimum $\chi^2$ is 52, with 52 degrees of freedom. 
The results are shown in Fig.~\ref{fig:2-c9_c9p}. The tension with the SM is still present in this set but only for $C_9$ and not for $C_9^{'}$ .

\subsection{Fit results for four operators}
\label{sec:4-op}
We expand here the study in the previous section by considering four operators in the fits, since there is a priori no reason that new physics should affect only two operators. We consider three possible sets including chirality flipped operators, and electron and muon contributions separately which are described in the following subsections.

\subsubsection{Fit results for \{$C_9, C_9^{'}, C_{10}, C_{10}^{'}$\}}
\begin{figure}[!h]
\begin{center}
\includegraphics[width=6.cm]{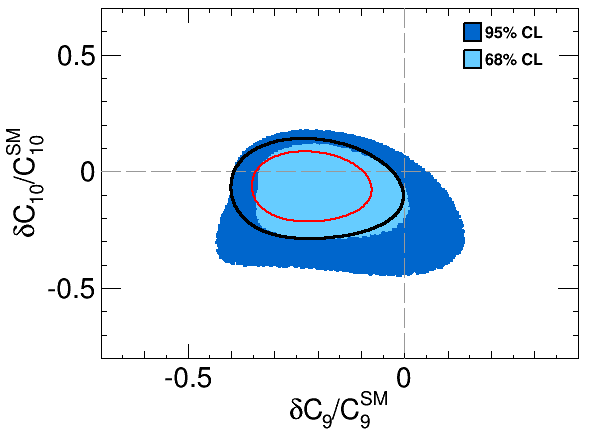}\quad\quad\includegraphics[width=6.cm]{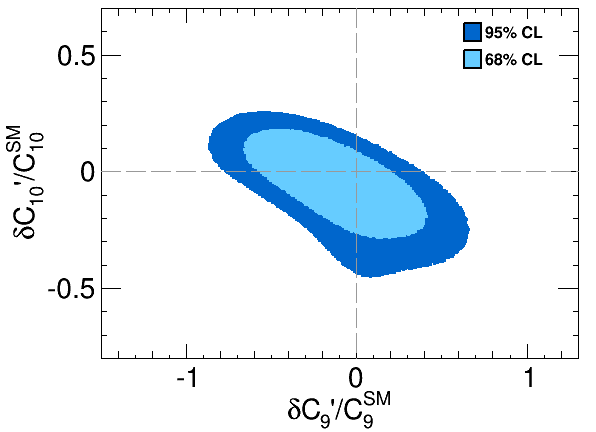}\\
\includegraphics[width=6.cm]{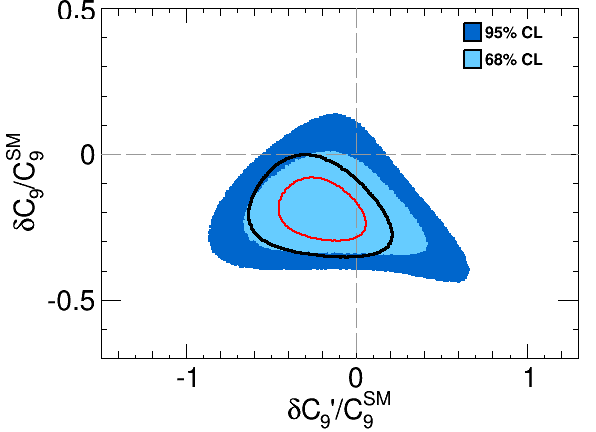}\quad\quad\includegraphics[width=6.cm]{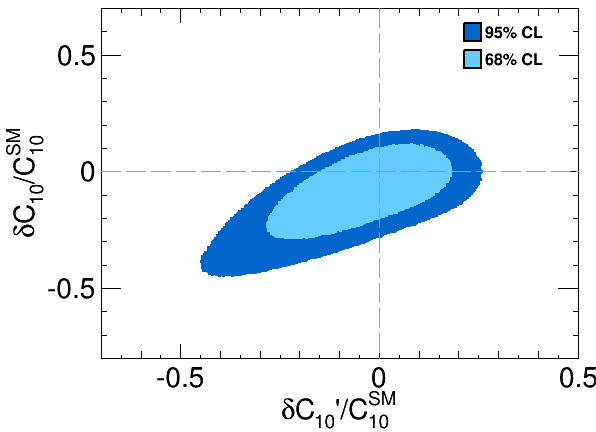}
\caption{Global fit results for $C_9, C_9^{'}, C_{10}, C_{10}^{'}$. The red and black contours in the upper left plot correspond to the 1 and 2$\sigma$ regions respectively, for the $(C_9,C_{10})$ fit presented in section~\ref{sec:2-op}. In the lower left plot the contours corresponding to 1 and 2$\sigma$ regions from the fit to ($C_9, C_9^{'}$) is superimposed.}
\label{fig:4a}
\end{center}
\end{figure}

The first set of four operators that we consider is an extension of \{$O_9, O_{10}$\} presented in section~\ref{sec:2-op} by adding also the chirality flipped operators \{$O_9^{'}, O_{10}^{'}$\}.
For this, we scan over $\delta C_9, \delta C_9^{'}, \delta C_{10}, \delta C_{10}^{'}$. 
In Fig.~\ref{fig:4a} we show the results of the global fit.
We see that in this scenario the agreement with the SM is better but only at the $2\sigma$ level. Negative contributions to $\delta C_9$ are more favoured, whereas all the other Wilson coefficients can keep their SM values.
To compare with the result of section~\ref{sec:2-op}, we superimpose the 1 and 2$\sigma$ contours from the fit to two operators only in Fig.~\ref{fig:4a}. This shows that considering only two operators leads to more restrictive results and some new physics contributions could be overlooked.

The set $\{C_9,C_9',C_{10},C'_{10}\}$ is an extension of both the $\{C_9,C_{10}\}$ and $\{C_9,C'_{9}\}$ sets, it is hence instructive to compare the $\chi^2$ values of the best fit-points. The best fit point of $\{C_9,C_9',C_{10},C'_{10}\}$ has a value of 51 (for 50 degrees of freedom), which is very similar to the values for the two other sets. Therefore, we can conclude that in the lepton universal scenario, adding the $C_{10}$ coefficients or the primed coefficients does not improve the fit.\\

\subsubsection{Fit results for \{$C_9^\mu, C_9^{'\mu}, C_9^e, C_9^{'e}$\}}
\begin{figure}[!h]
\begin{center}
\includegraphics[width=6.cm]{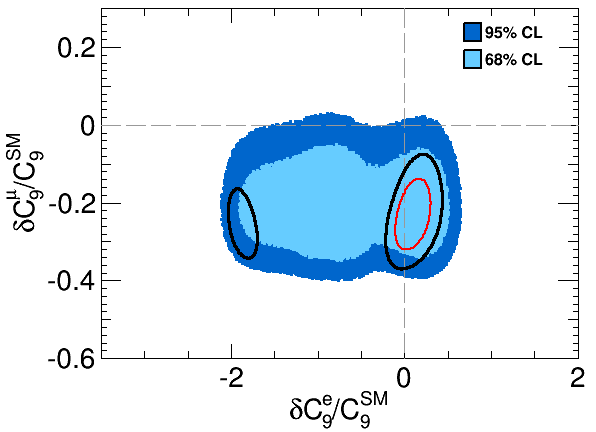}\quad\quad\includegraphics[width=6.cm]{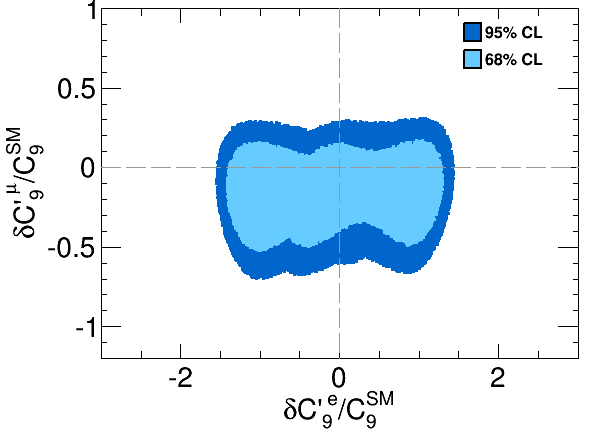}\\
\includegraphics[width=6.cm]{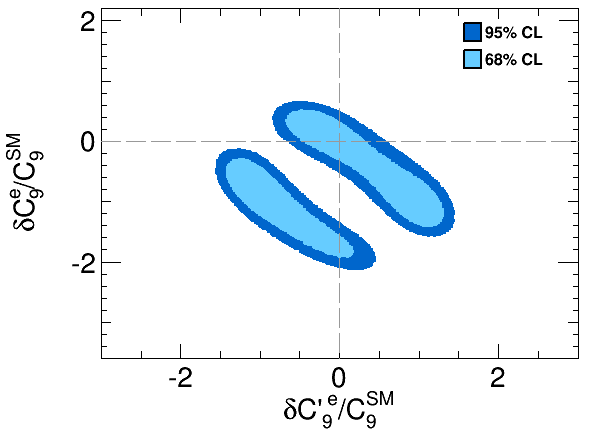}\quad\quad\includegraphics[width=6.cm]{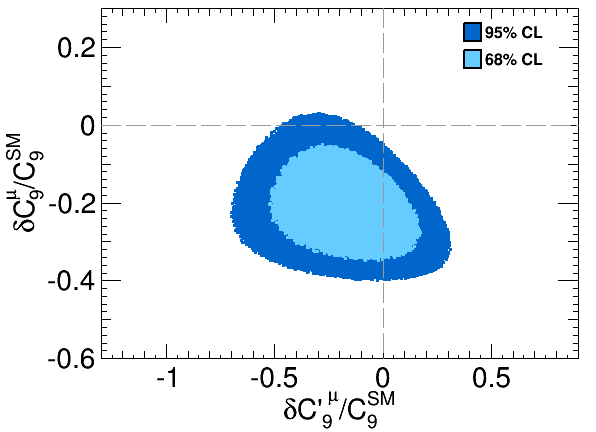}
\caption{Global fit results for $C_9^\mu, C_9^{'\mu}, C_9^e, C_9^{'e}$. The red and black  contours in the upper left plot correspond to the 1 and 2$\sigma$ regions respectively, for the $(C^e_9,C^\mu_9)$ fit presented in section~\ref{sec:2-op}.}
\label{fig:4c}
\end{center}
\end{figure}

We expand here the study in section \ref{sec:2-op} by adding to the \{$O_9^\mu, O_9^e$\} set the chirality flipped operators, namely \{$O_9^{'\mu}, O_9^{'e}$\}.
We scan therefore over $\delta C_9^{\mu}, \delta C_9^{e}, \delta C_9^{'\mu}, \delta C_9^{'e}$. 
The results are shown in Fig.~\ref{fig:4c}.

In order to compare the results with the scan for two operators \{$O_9^\mu, O_9^e$\} only, we overlay the contours corresponding to the 1 and 2$\sigma$ fit result for \{$C_9^\mu, C_9^e$\} in Fig.~\ref{fig:4c} in the plane ($\delta C_9^\mu, \delta C_9^e$).  
The comparison with the results obtained in the \{$C^\mu_9,C^e_9$\} case clearly shows that considering only the modification of two Wilson coefficients leads to much more restrictive results.

We see that the SM is disfavoured at the 2$\sigma$ level, even if the agreement is now improved compared to the two operator case. In particular, it is now possible to have simultaneously $\delta C^e_9=0$ and $\delta C^\mu_9=0$ at the 2$\sigma$ level. Yet there is still tension in the muon sector for $C_9$. However, we emphasize also that sizeable new physics contributions to the other three operators are allowed at the 1$\sigma$ level as it is visible in the plots of Fig.~\ref{fig:4c}.

The set $\{C_9^\mu,C_9^{'\mu},C_9^e,C_9^{'e}\}$ is a direct extension of $\{C_9^\mu,C_9^e\}$. Its best fit point has a $\chi^2$ of 42, therefore adding the primed coefficients only slightly improves the fit. If we now compare $\{C_9^\mu,C_9^{'\mu},C_9^e,C_9^{'e}\}$ to the $\{C_9,C^{'}_{9}\}$ set (which can be obtained by setting $\delta C_9^\mu=\delta C_9^e$ and $\delta C^{'\mu}_9=\delta C^{'e}_9$), we notice an improvement of about 2.6$\sigma$, which shows that considering non-universal lepton couplings improves the fit. The rigorous statement is the following: Assuming that the four operator scenario with $\{C_9^\mu,C_9^{'\mu},C_9^e,C_9^{'e}\}$ is correct, then the one with the two operators with $\{C_9,C^{'}_{9}\}$ is ruled out at 2.6$\sigma$. This allows at least for a qualitative comparison of the various four operator fits.\\

\subsubsection{Fit results for \{$C_9^\mu, C_9^e, C_{10}^\mu, C_{10}^e$\}}
\begin{figure}[!h]
\begin{center}
\includegraphics[width=6.cm]{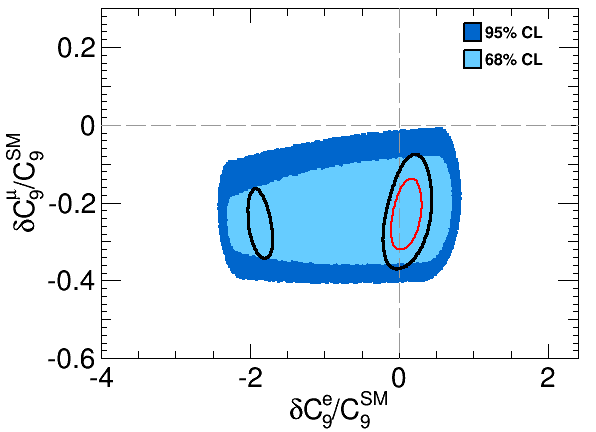}\quad\quad\includegraphics[width=6.cm]{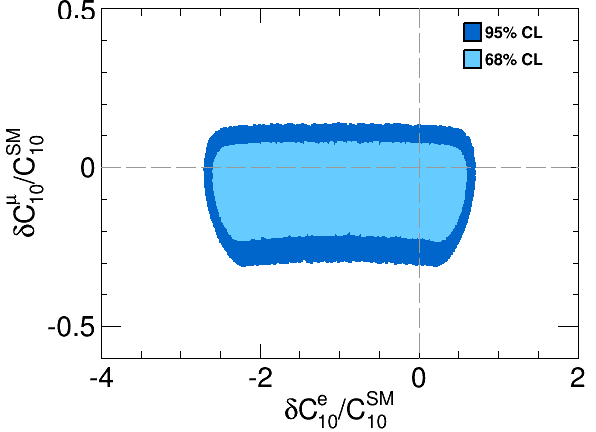}\\
\includegraphics[width=6.cm]{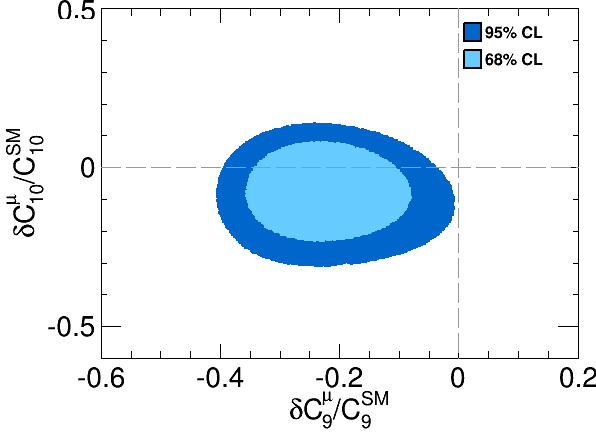}\quad\quad\includegraphics[width=6.cm]{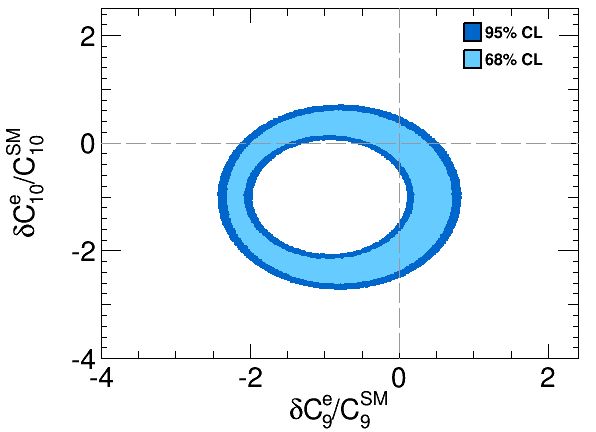}
\caption{Global fit results for $C_9^\mu, C_9^e, C_{10}^\mu, C_{10}^e$. The red and black contours correspond to the 1 and 2$\sigma$ regions respectively of the $(C^e_9,C^\mu_9)$ fit presented in section~\ref{sec:2-op}.}
\label{fig:4b}
\end{center}
\end{figure}

We consider finally the set with both $O_9$ and $O_{10}$ but allow for lepton non-universality by considering separately the electron and muon contributions, neglecting the chirality flipped operators. 
We therefore perform a scan over $\delta C_9^{\mu}, \delta C_9^{e}, \delta C_{10}^{\mu}, \delta C_{10}^{e}$. The results are displayed in Fig.~\ref{fig:4b}.

As in the other cases, having all the Wilson coefficients at their SM values is disfavoured at the 2$\sigma$ level, yet allowing for a negative contribution to $C^\mu_9$ sorts this problem. However, as the plots illustrate, we note again that also large new physics contributions to $C_9^e$, $C_{10}^\mu$ and $C_{10}^e$ are allowed within the 1$\sigma$ level.

The comparison with the results obtained in the $\{C^\mu_9,C^e_9\}$ case clearly shows that considering only the modification of two Wilson coefficients leads to much restrictive results and overlooks other viable possibilities for new physics contributions.

The set $\{C_9^\mu,C_9^e,C_{10}^\mu,C_{10}^e\}$ is a direct extension of $\{C_9^\mu,C_9^e\}$. The best fit point in that scenario has a $\chi^2$ of 43. So the addition of the $C_{10}^{\mu,e}$ coefficients only slightly improves the fit. If we compare it to the $\{C_9,C_{10}\}$ set, we see an improvement of 2.5$\sigma$, again favouring the non-universal extension against the universal one.

\subsection{Fit results assuming left-handed leptons}

Finally, we make the assumption that we have left-handed leptons only, which represents an attractive option in model building beyond the SM. In this case one finds the following relations between the Wilson coefficients:
\begin{equation}
\delta C_{LL}^{i}\equiv \delta C_9^{i} =-\delta C_{10}^{i}\,,\,\,\,\,\,  \delta C_{RL}^i \equiv \delta C^{\prime\,i}_9 =-\delta C^{\prime i}_{10}\,.
\end{equation}
Here we introduced the quantities $C_{XY}^i$ where $i$ is the flavour index, $X$ denotes the chirality of the quark current and $Y$ of the lepton one.\footnote{For completeness, we note that assuming right-handed leptons only, one gets the following relations: 
\begin{equation}
\delta C_{RR}^i \equiv \delta C^{\prime\,i}_9 = \delta C^{\prime\,i}_{10}\,,\,\,\,   \delta C_{LR}^i \equiv \delta{C^{i}_9} = \delta C^{i}_{10}\,.
\end{equation}
}

\begin{figure}[!h]
\begin{center}
\includegraphics[width=6.cm]{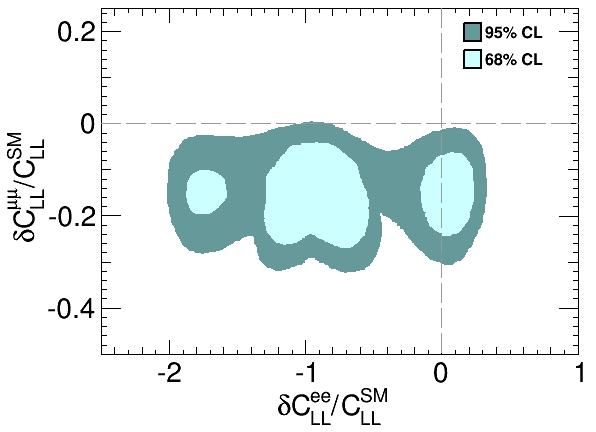}\quad\quad\includegraphics[width=6.cm]{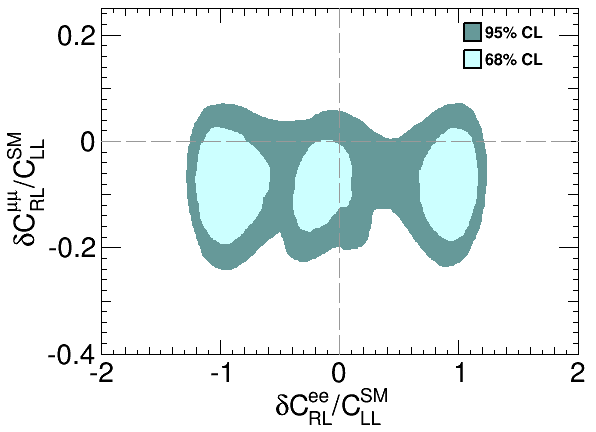}
\caption{Global fit results for $C_{LL}^{\mu\mu}$, $C_{LL}^{ee}$, $C_{RL}^{\mu\mu}$, $C_{RL}^{ee}$.}
\label{fig:4-GI}
\end{center}
\end{figure}

We perform a fit for $C_{LL}^{\mu\mu}$, $C_{LL}^{ee}$, $C_{RL}^{\mu\mu}$, $C_{RL}^{ee}$.
The results are shown in Fig.~\ref{fig:4-GI}. The fit shows a large 2$\sigma$ region and three separate 1$\sigma$ areas. The coefficient $C_{LL}^{\mu\mu}$ deviates slightly from its SM value at the 2$\sigma$ level, while the other coefficients are compatible with their SM values. Both $C_{LL}^{\mu\mu}$ and $C_{RL}^{\mu\mu}$ can vary by only 30\% away from the SM value, while the electron operators can have a larger deviation. 
The minimum $\chi^2$ of the fit is 41.

\begin{figure}[!h]
\begin{center}
\includegraphics[width=6.cm]{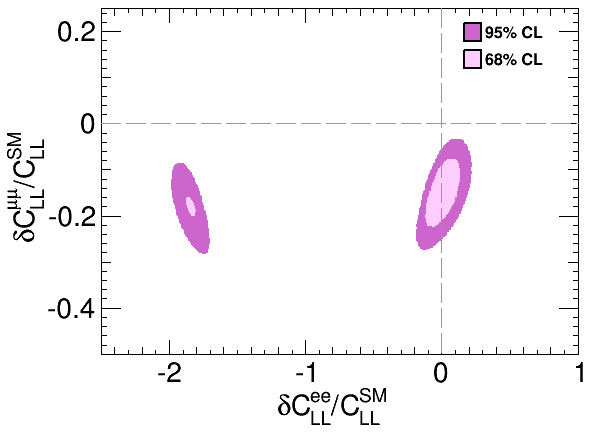}
\caption{Global fit results for $C_{LL}^{\mu\mu}$, $C_{LL}^{ee}$.}
\label{fig:2-GI}
\end{center}
\end{figure}

We also made a fit with two coefficients $C_{LL}^{\mu\mu}$, $C_{LL}^{ee}$, setting the RL operators to zero. The result is displayed in Fig.~\ref{fig:2-GI}. As seen in section~\ref{sec:4-op}, the resulting fit regions are much smaller, which shows that it is important to consider the possibility of RL new physics operators as well. In this case, the minimum $\chi^2$ is 46, showing that assuming the four operator set is the correct one, the two operator set is disfavoured by about 1.5$\sigma$.

\section{Comparison of exclusive and inclusive $b \to s \ell\ell$ observables}

The inclusive mode  $B\to X_s \ell^+ \ell^-$ can only be measured at $e^+ e^-$ machines and is theoretically cleaner than the exclusive modes~\cite{Hurth:2010tk,Hurth:2012vp}. 
The theoretical accuracy in the low-$q^2$  region is of the order of $10\%$~\cite{Huber:2007vv}. But the branching fraction has been measured by Belle and BaBar using the sum-of-exclusive technique only. The latest published measurement  of Belle~\cite{Iwasaki:2005sy}  is based on a sample of $152 \times 10^6$ $B \bar B$ events only, which corresponds to less than $30\%$ of the dataset available at the end of the Belle experiment.
BaBar has just recently presented an analysis based on the whole dataset using a sample of $471 \times 10^6$ $B \bar B$ events~\cite{Lees:2013nxa} overwriting 
the previous measurement from 2004 based on $89 \times 10^6 $ $B \bar B$  events~\cite{Aubert:2004it}.

In order to compare different sets of observables, we consider the operators $O_7$, $O_9$ and $O_{10}$ which are the most relevant operators for the semileptonic $B$ decays. Again we use the $\Delta \chi^2$ fit method to obtain the exclusion plots of the Wilson coefficients.\footnote{We note that the $\chi^2$ method  is not suitable for this kind of comparison 
because the exclusion plots would change if some less sensitive observables were removed from the fit. However, we checked first that the $\chi^2$ method signals the overall consistency of the separate fits.}

\begin{figure}[!h]
\begin{center}
\includegraphics[width=6.cm]{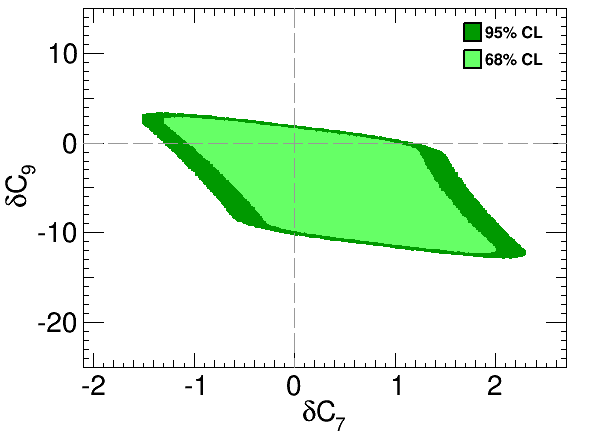}\quad\quad
\includegraphics[width=6.cm]{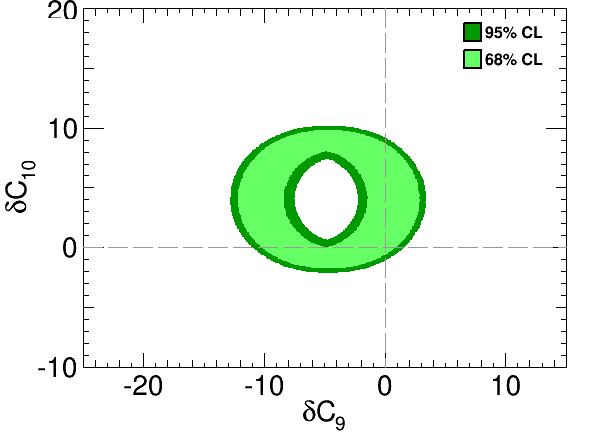}\\
\includegraphics[width=6.cm]{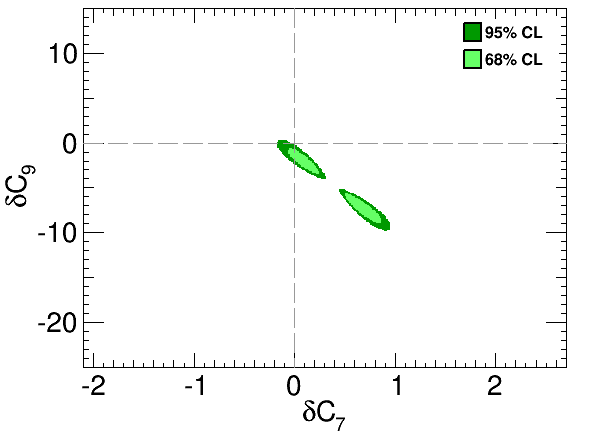}\quad\quad
\includegraphics[width=6.cm]{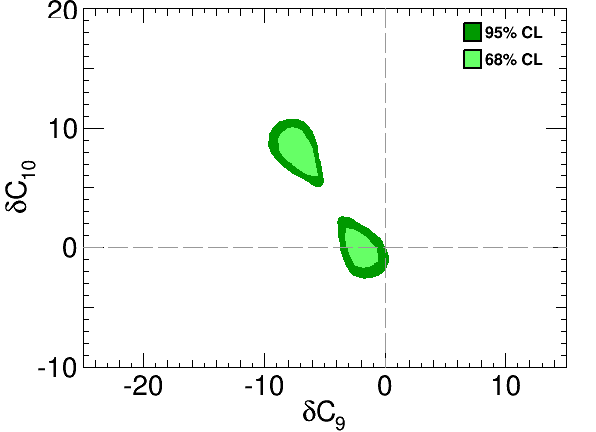}
\includegraphics[width=6.cm]{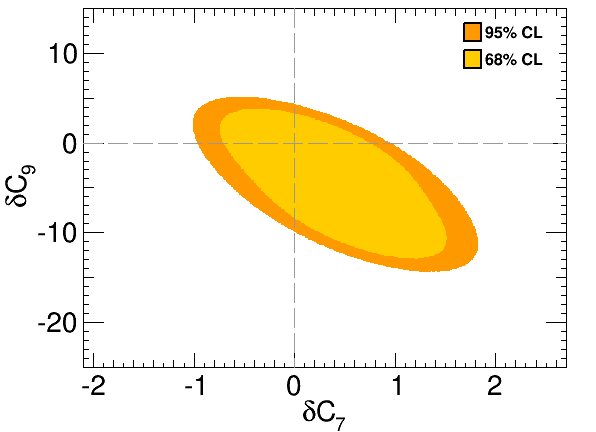}\quad\quad
\includegraphics[width=6.cm]{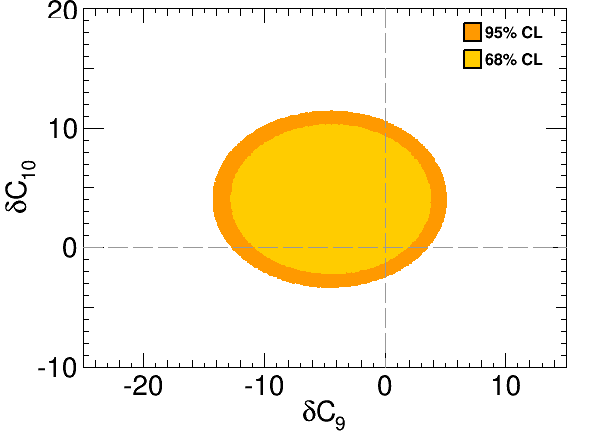}
\caption{Fit results for the new physics contributions to $C_7$, $C_9$ and $C_{10}$, using only  BR$(B \to K^0 \mu^+ \mu^-)$, BR$(B^+ \to K^+ \mu^+ \mu^-)$, $R_K$ (first row),  using only $B\to K^* \mu^+\mu^-$ observables (second row), and using the current measurements of BR($B\to X_s \mu^+\mu^-$) at low- and high-$q^2$ (third row).\label{fig:bslla}}
\end{center}
\end{figure}
First we make  the global fit using only $B\to X_s  \ell^+\ell^-$ branching ratios in the low-$q^2$ and high-$q^2$ regions with $\ell = \mu, e$. We note that there is no sign of lepton non-universality in the published data. We find that the $\chi^2$ fit 
using the two branching ratios with $\ell = e$ only is not very good; there is no compatibility at $68\%$ C.L. , so for the $\Delta \chi^2$-metrology we restrict ourselves to the 
case $\ell= \mu$. We compare  this with two other fits using exclusive observables, one with BR$(B \to K^0 \mu^+ \mu^-)$, BR$(B^+ \to K^+ \mu^+ \mu^-)$, $R_K$,
and one with all $B \to K^* \mu^+\mu^-$ observables.
In Fig.~\ref{fig:bslla}, we illustrate the results of the $\Delta \chi^2$ fit for the relevant Wilson coefficients. The upper row shows the fit based on the exclusive observables 
BR$(B \to K^0 \mu^+ \mu^-)$, BR$(B^+ \to K^+ \mu^+ \mu^-)$, $R_K$; the middle row shows fit based on $B \rightarrow K^* \ell^+\ell^-$ observables; and the lower row the one based on the measurements of the inclusive ($B \to X_s\mu^+\mu^-$) branching ratio in the low- and high-$q^2$ regions. It is remarkable that all three sets of exclusion plots are nicely compatible with each other.  This is a non-trivial consistency check. 

At the moment, the measurements of the $B \rightarrow K^* \ell^+\ell^-$ are the most powerful ones. 
However, the final word of Belle is still pending and, moreover,  there will be a Super-$B$ factory Belle-II with a final integrated luminosity of 50 ab$^{-1}$~\cite{belle2}.
There is a recent analysis~\cite{Kevin2} of the expected total uncertainty on the partial decay width and the forward-backward asymmetry in several bins of dilepton mass-squared for the fully inclusive $B \to  X_s \ell^+\ell^-$ decays assuming a 50 ab$^{-1}$ total integrated luminosity. Based on some reasonable 
assumptions (for details see Ref.~\cite{Hurth:2013ssa}), one finds a relative fractional uncertainty of $2.9\%$  ($4.1\%$) for the branching fraction in the low- (high-) $q^2$ region and a total absolute uncertainty of 0.050 in the low-$q^2$ bin 1 ($1<q^2<3.5$ GeV$^2$), 0.054 in the low-$q^2$ bin 2 ($3.5<q^2<6$ GeV$^2$) and 0.058 in the high-$q^2$ interval ($q^2>14.4$ GeV$^2$) for the {\it normalised} $A_{FB}$.
Hence, the inclusive mode will lead to very strong constraints on the Wilson coefficients and to a more significant cross-check of the new physics hypothesis. 

\begin{figure}[t!]
\begin{center}
\includegraphics[width=8.cm]{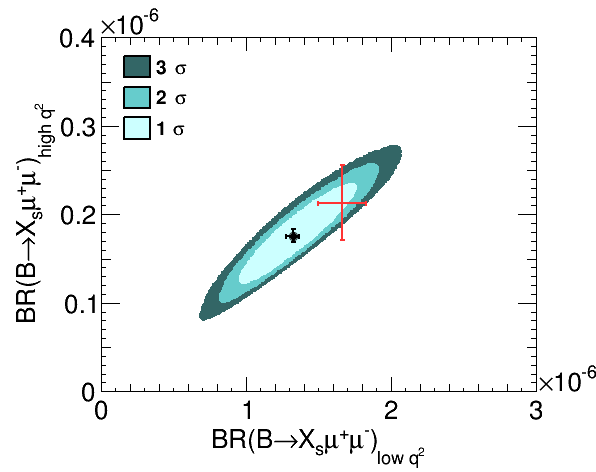}
\caption{1, 2 and 3$\sigma$ predictions for the branching ratio at low- and high-$q^2$ within the model-independent analysis. 
Future measurement at the high-luminosity Belle-II Super-$B$-Factory assuming the best-fit point of the model-independent analysis as central value (black) and the SM predictions (red/grey).\label{fig:bsllc}}
\end{center}
\end{figure}
\begin{figure}[t!]
\begin{center}
\includegraphics[width=8.cm]{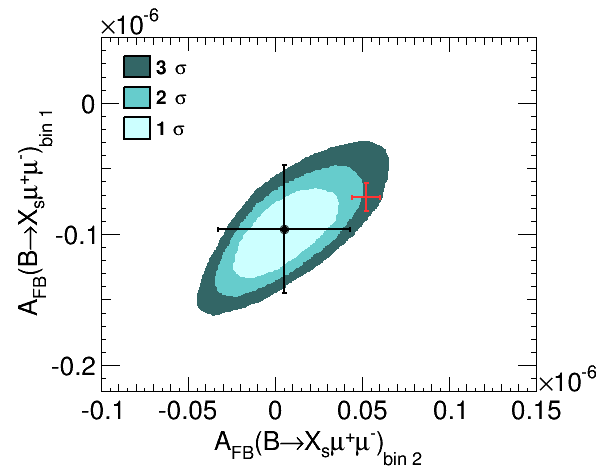}
\caption{1, 2 and 3$\sigma$ predictions for the {\it unnormalised} forward-backward asymmetry in bin 1 ($1<q^2<3.5$ GeV$^2$) and in bin 2 ($3.5<q^2<6$ GeV$^2$) within the model-independent analysis. Future measurement at the high-luminosity Belle-II Super-$B$-Factory assuming the best-fit point of the 
model-independent analysis as central value (black) and the SM predictions (red/grey).\label{fig:bslld}}
\end{center}
\end{figure}
We illustrate the usefulness of these  future measurements of the inclusive mode at Belle-II in the following way: 
We make a model independent fit for the coefficients $C_7$, $C_8$, $C_9$, $C_{10}$ and $C_{l}$ (for notation see Ref.~\cite{Hurth:2013ssa}). In addition to the observables given in Tables~\ref{tab:BtoKstar} and \ref{tab:rest}, we consider the inclusive branching ratio of $B\to X_s \gamma$ as well as the isospin asymmetry in $B\to K^* \gamma$ decay which are relevant to constrain $C_7$ and $C_8$.  
Based on our model-independent analysis we predict the branching ratios at low- and high-$q^2$. In Fig.~\ref{fig:bsllc},
we show the 1, 2, and 3$\sigma$ ranges for these observables. In addition, we add the future measurements at Belle-II assuming the best fit solution of our  model-independent analysis as central value. These measurements are indicated by the black error bars. They should be compared with the theoretical SM predictions given by the red (grey) error bars. 
Fig.~\ref{fig:bsllc} indicates that the future measurement of the inclusive branching ratios separates nicely the SM prediction and the model-independent best fit point. Moreover, the future measurement of the forward-backward asymmetry at Belle-II will allow to separate the potential new physics measurement from the SM prediction in a significant way as shown in Fig.~\ref{fig:bslld}.

\section{Conclusions}

We present here for the first time global fits to the complete $b\to s\ell\ell$ dataset available from $B$ factories and from LHC, in particular addressing the two observed tensions in the angular analysis of the exclusive decay $B \to K^* \mu^+\mu^-$ and the ratio $R_K$. We perform several global fits using different sets of two or four vector and axial vector operators allowing for non-universality. 

Comparing the 4-operator and 2-operator fit results we have shown that
while considering 2-operator fits can be illustrative for where in the
parameter space new physics could be found it can be
too restrictive and maybe even misleading since a large area of new physics
parameter space might be unjustifiably overlooked.

Considering the full set of 8 semileptonic operators $\{C_9^\mu,C_9^{'\mu},C_9^e,C_9^{'e},C_{10}^\mu,C_{10}^{'\mu},C_{10}^e,C_{10}^{'e}\}$ and comparing the minimum $\chi^2$ of the different subsets can lead to a determination of the most relevant operators. This comparison reveals that the sets with lepton non-universality, namely $\{C_9^\mu,C_9^{'\mu},C_9^e,C_9^{'e}\}$ and \{$C_9^\mu, C_9^e, C_{10}^\mu, C_{10}^e$\} give the best fit to the data.  

Our global fits show that simultaneous agreement of all the Wilson coefficients with the Standard Model is ruled out by 2$\sigma$ at least. This problem can be resolved if the new physics contribution to $C_9^\mu$ is negative. However, we emphasize that sizeable new physics contributions to the other operators are allowed at the 1$\sigma$ level.

If these tensions are not resolved in the near future, we have demonstrated that the future measurements of the inclusive $b\to s\ell\ell$ observables by Belle II will allow for a powerful cross-check. The present data on inclusive and exclusive decays are nicely compatible with each other and there is no sign of lepton non-universality in the published data on the inclusive mode.

\section*{Acknowledgement} 

TH thanks the CERN theory group for its hospitality during his regular visits to CERN where part of this work was written. SN thanks the department of Physics of Isfahan University of Technology for their hospitality where part of this work was done. The authors are grateful to Andreas Crivellin for many useful discussions and to Nicola Serra for the help with the experimental correlations for the $B \rightarrow K^{*} \mu^+\mu^-$ angular observables, and to J\'er\^ome Charles for his valuable comments. 



\begin{thebibliography}{12}



\bibitem{Aaij:2013qta}
  R.~Aaij {\it et al.}  [LHCb Collaboration],
  Phys.\ Rev.\ Lett.\  {\bf 111} (2013) 191801
  [arXiv:1308.1707 [hep-ex]].

\bibitem{Jager:2012uw}
  S.~J\"ager and J.~Martin Camalich,
  JHEP {\bf 1305} (2013) 043
  [arXiv:1212.2263 [hep-ph]].
  
\bibitem{Descotes-Genon:2013wba} 
  S.~Descotes-Genon, J.~Matias and J.~Virto,
  Phys.\  Rev.\  D {\bf 88} (2013) 074002
  [arXiv:1307.5683 [hep-ph]].

\bibitem{Altmannshofer:2013foa} 
  W.~Altmannshofer and D.~M.~Straub,
  Eur.\ Phys.\ J.\ C {\bf 73} (2013) 2646
  [arXiv:1308.1501 [hep-ph]].

\bibitem{Gauld:2013qba}
  R.~Gauld, F.~Goertz and U.~Haisch,
  Phys.\ Rev.\ D {\bf 89} (2014) 015005
  [arXiv:1308.1959 [hep-ph]].

\bibitem{Hambrock:2013zya}
  C.~Hambrock, G.~Hiller, S.~Schacht and R.~Zwicky,
  Phys.\ Rev.\ D {\bf 89} (2014) 074014
  [arXiv:1308.4379 [hep-ph]].

\bibitem{Buras:2013qja}
  A.~J.~Buras and J.~Girrbach,
  JHEP {\bf 1312} (2013) 009
  [arXiv:1309.2466 [hep-ph]].

\bibitem{Gauld:2013qja}
  R.~Gauld, F.~Goertz and U.~Haisch,
  JHEP {\bf 1401} (2014) 069
  [arXiv:1310.1082 [hep-ph]].

\bibitem{Datta:2013kja}
  A.~Datta, M.~Duraisamy and D.~Ghosh,
  Phys.\ Rev.\ D {\bf 89} (2014) 071501
  [arXiv:1310.1937 [hep-ph]].

\bibitem{Beaujean:2013soa}
  F.~Beaujean, C.~Bobeth and D.~van Dyk,
  Eur.\ Phys.\ J.\ C {\bf 74} (2014) 2897
  [arXiv:1310.2478 [hep-ph]].

\bibitem{Horgan:2013pva}
  R.~R.~Horgan, Z.~Liu, S.~Meinel and M.~Wingate,
  Phys.\ Rev.\ Lett.\  {\bf 112} (2014) 212003
  [arXiv:1310.3887 [hep-ph]].

\bibitem{Buras:2013dea}
  A.~J.~Buras, F.~De Fazio and J.~Girrbach,
  JHEP {\bf 1402} (2014) 112
  [arXiv:1311.6729 [hep-ph]].

\bibitem{Hurth:2013ssa}
  T.~Hurth and F.~Mahmoudi,
  JHEP {\bf 1404} (2014) 097
  [arXiv:1312.5267 [hep-ph]].

\bibitem{Descotes-Genon:2014uoa}
  S.~Descotes-Genon, L.~Hofer, J.~Matias and J.~Virto,
  arXiv:1407.8526 [hep-ph].

\bibitem{Khodjamirian:2010vf}
  A.~Khodjamirian, T.~Mannel, A.~A.~Pivovarov and Y.~-M.~Wang,
  JHEP {\bf 1009} (2010) 089
  [arXiv:1006.4945 [hep-ph]].

\bibitem{Lyon:2014hpa}
  J.~Lyon and R.~Zwicky,
  arXiv:1406.0566 [hep-ph].

\bibitem{Aaij:2014ora}
  R.~Aaij {\it et al.}  [LHCb Collaboration],
  Phys.\ Rev.\ Lett.\  {\bf 113} (2014) 151601
  [arXiv:1406.6482 [hep-ex]].

\bibitem{Alonso:2014csa}
  R.~Alonso, B.~Grinstein and J.~M.~Camalich,
  arXiv:1407.7044 [hep-ph].

\bibitem{Hiller:2014yaa}
  G.~Hiller and M.~Schmaltz,
  Phys.\ Rev.\ D {\bf 90} (2014) 054014
  [arXiv:1408.1627 [hep-ph]].

\bibitem{Ghosh:2014awa}
  D.~Ghosh, M.~Nardecchia and S.~A.~Renner,
  arXiv:1408.4097 [hep-ph].

\bibitem{Biswas:2014gga}
  S.~Biswas, D.~Chowdhury, S.~Han and S.~J.~Lee,
  arXiv:1409.0882 [hep-ph].
  
\bibitem{Straub} 
See also talk by D. Straub at the ``Implications of LHCb measurements and future prospects'' workshop, CERN, October 16th 2014.   
  
\bibitem{Hurth:2007xa}
  T.~Hurth,
  Int.\ J.\ Mod.\ Phys.\ A {\bf 22} (2007) 1781
  [hep-ph/0703226].

\bibitem{Hurth:2003vb}
  T.~Hurth,
  Rev.\ Mod.\ Phys.\  {\bf 75} (2003) 1159
  [hep-ph/0212304].
  
\bibitem{Hiller:2003js}
  G.~Hiller and F.~Kruger,
  Phys.\ Rev.\ D {\bf 69} (2004) 074020
  [hep-ph/0310219].

\bibitem{Beneke:2001at}
  M.~Beneke, T.~Feldmann and D.~Seidel,
  Nucl.\ Phys.\ B {\bf 612} (2001) 25
  [hep-ph/0106067].

\bibitem{Beneke:2004dp}
  M.~Beneke, T.~Feldmann and D.~Seidel,
  Eur.\ Phys.\ J.\ C {\bf 41} (2005) 173
  [hep-ph/0412400].
  
\bibitem{Charles:1998dr}
  J.~Charles, A.~Le Yaouanc, L.~Oliver, O.~Pene and J.~C.~Raynal,
  Phys.\ Rev.\ D {\bf 60} (1999) 014001
  [hep-ph/9812358].

\bibitem{Beneke:2000wa}
  M.~Beneke and T.~Feldmann,
  Nucl.\ Phys.\  B {\bf 592}, 3 (2001)
  [arXiv:hep-ph/0008255].

\bibitem{Bobeth:2007dw}
  C.~Bobeth, G.~Hiller and G.~Piranishvili,
  JHEP {\bf 0712} (2007) 040
  [arXiv:0709.4174 [hep-ph]].

\bibitem{Bobeth:2011nj}
  C.~Bobeth, G.~Hiller, D.~van Dyk and C.~Wacker,
  JHEP {\bf 1201} (2012) 107
  [arXiv:1111.2558 [hep-ph]].
  
\bibitem{Bouchard:2013eph}
  C.~Bouchard, G.~P.~Lepage, C.~Monahan, H.~Na and J.~Shigemitsu,
  Phys.\ Rev.\ D {\bf 88} (2013) 054509
  [arXiv:1306.2384 [hep-lat]].

\bibitem{Horgan:2013hoa}
  R.~R.~Horgan, Z.~Liu, S.~Meinel and M.~Wingate,
  Phys.\ Rev.\ D {\bf 89} (2014) 094501
  [arXiv:1310.3722 [hep-lat]].

\bibitem{Egede:2008uy}
  U.~Egede, T.~Hurth, J.~Matias, M.~Ramon and W.~Reece,
  JHEP {\bf 0811} (2008) 032
  [arXiv:0807.2589 [hep-ph]].

\bibitem{Egede:2010zc}
  U.~Egede, T.~Hurth, J.~Matias, M.~Ramon and W.~Reece,
  JHEP {\bf 1010} (2010) 056
  [arXiv:1005.0571 [hep-ph]].

\bibitem{Descotes-Genon:2013vna} 
  S.~Descotes-Genon, T.~Hurth, J.~Matias and J.~Virto,
  JHEP {\bf 1305} (2013) 137
  [arXiv:1303.5794 [hep-ph]].

\bibitem{Hermann:2013kca} 
  T.~Hermann, M.~Misiak and M.~Steinhauser,
  JHEP {\bf 1312} (2013) 097
  [arXiv:1311.1347 [hep-ph]].

\bibitem{Bobeth:2013tba} 
  C.~Bobeth, M.~Gorbahn and E.~Stamou,
  Phys. Rev. D{\bf 89} (2014) 034023
  [arXiv:1311.1348 [hep-ph]].

\bibitem{Aaij:2013hha}
  R.~Aaij {\it et al.}  [LHCb Collaboration],
  JHEP {\bf 1305} (2013) 159
  [arXiv:1304.3035 [hep-ex]].

\bibitem{Lees:2013nxa}
  J.~P.~Lees {\it et al.}  [BaBar Collaboration],
  Phys.\ Rev.\ Lett.\  {\bf 112} (2014) 211802
  [arXiv:1312.5364 [hep-ex]].

\bibitem{Agashe:2014kda}
  K.~A.~Olive {\it et al.}  [Particle Data Group Collaboration],
  Chin.\ Phys.\ C {\bf 38} (2014) 090001.
  
\bibitem{Aoki:2013ldr}
  S.~Aoki, Y.~Aoki, C.~Bernard, T.~Blum, G.~Colangelo, M.~Della Morte, S.~DÃ¼rr and A.~X.~El Khadra {\it et al.},
  Eur.\ Phys.\ J.\ C {\bf 74} (2014) 9,  2890
  [arXiv:1310.8555 [hep-lat]].

\bibitem{Ball:2006fz}
  P.~Ball and R.~Zwicky,
  JHEP {\bf 0602} (2006) 034
  [hep-ph/0601086].

\bibitem{Ball:2006wn}
  P.~Ball, V.~M.~Braun and A.~Lenz,
  JHEP {\bf 0605} (2006) 004
  [hep-ph/0603063].

\bibitem{Bazavov:2009bb}
  A.~Bazavov, D.~Toussaint, C.~Bernard {\it et al.},
  Rev.\ Mod.\ Phys.\  {\bf 82} (2010) 1349
  [arXiv:0903.3598 [hep-lat]].

\bibitem{Mahmoudi:2014mja}
  F.~Mahmoudi, S.~Neshatpour and J.~Virto,
  Eur.\ Phys.\ J.\ C {\bf 74} (2014) 2927
  [arXiv:1401.2145 [hep-ph]].
  
\bibitem{Mahmoudi:2012un}
  F.~Mahmoudi, S.~Neshatpour and J.~Orloff,
  JHEP {\bf 1208} (2012) 092
  [arXiv:1205.1845 [hep-ph]].

\bibitem{Aaij:2013iag}
R. Aaij {\it et al.} [LHCb Collaboration], 
JHEP {\bf 1308} (2013) 131 [arXiv:1304.6325 [hep-ex]]

\bibitem{Aaij:2013aka}
  R. Aaij {\it et al.}  [LHCb Collaboration],
  Phys.\ Rev.\ Lett.\  {\bf 111} (2013) 101805
  [arXiv:1307.5024 [hep-ex]].

\bibitem{Chatrchyan:2013bka}
  S.~Chatrchyan {\it et al.}  [CMS Collaboration],
  Phys.\ Rev.\ Lett.\  {\bf 111} (2013) 101804
  [arXiv:1307.5025 [hep-ex]].

\bibitem{CMSandLHCbCollaborations:2013pla}
  [CMS and LHCb Collaborations],
  CMS-PAS-BPH-13-007, LHCb-CONF-2013-012.
  
\bibitem{Aaij:2014pli}
  R.~Aaij {\it et al.}  [LHCb Collaboration],
  JHEP {\bf 1406} (2014) 133
  [arXiv:1403.8044 [hep-ex]].

\bibitem{Mahmoudi:2007vz}
  F.~Mahmoudi,
  Comput.\ Phys.\ Commun.\  {\bf 178} (2008) 745
  [arXiv:0710.2067 [hep-ph]].

\bibitem{Mahmoudi:2008tp}
  F.~Mahmoudi,
  Comput.\ Phys.\ Commun.\  {\bf 180} (2009) 1579
  [arXiv:0808.3144 [hep-ph]].

\bibitem{Hurth:2012jn}
  T.~Hurth and F.~Mahmoudi,
  Nucl.\ Phys.\ B {\bf 865} (2012) 461
  [arXiv:1207.0688 [hep-ph]].

\bibitem{Hurth:2010tk} 
  T.~Hurth and M.~Nakao,
  Ann.\ Rev.\ Nucl.\ Part.\ Sci.\  {\bf 60} (2010) 645
  [arXiv:1005.1224 [hep-ph]].

\bibitem{Hurth:2012vp}
  T.~Hurth and F.~Mahmoudi,
  Rev.\ Mod.\ Phys.\  {\bf 85} (2013) 795
  [arXiv:1211.6453 [hep-ph]].

\bibitem{Huber:2007vv}
  T.~Huber, T.~Hurth and E.~Lunghi,
  Nucl.\ Phys.\ B {\bf 802} (2008) 40
  [arXiv:0712.3009 [hep-ph]].

\bibitem{Iwasaki:2005sy}
  M.~Iwasaki {\it et al.}  [Belle Collaboration],
  Phys.\ Rev.\ D {\bf 72} (2005) 092005
  [hep-ex/0503044].

\bibitem{Aubert:2004it}
  B.~Aubert {\it et al.}  [BABAR Collaboration],
  Phys.\ Rev.\ Lett.\  {\bf 93} (2004) 081802
  [hep-ex/0404006].

\bibitem{belle2}
http://belle2.kek.jp/

\bibitem{Kevin2}
Kevin Flood, private communication.


  
\end{thebibliography}
\end{document}